\documentclass[aps,twocolumn,epsf,showpacs]{revtex4} 
\usepackage{graphicx}
 
\begin{document}
   
 

\title{Bright solitons in coupled defocusing NLS equation 
supported by coupling: Application to Bose-Einstein
Condensation\footnote{{\it E-mail address:}
adhikari@ift.unesp.br (S.K. Adhikari).}}
 
\author{Sadhan K. Adhikari}
\affiliation
{Instituto de F\'{\i}sica Te\'orica, Universidade Estadual
Paulista, 01.405-900 S\~ao Paulo, S\~ao Paulo, Brazil\\}

\date{\today}
 
 
\begin{abstract}

We demonstrate the formation of bright solitons in coupled defocusing
nonlinear Schr\"odinger (NLS) equation supported by attractive coupling.  
As an application we use a time-dependent dynamical mean-field model to
study the formation of stable bright solitons in two-component repulsive
Bose-Einstein condensates (BECs) supported by interspecies attraction in a
quasi one-dimensional geometry. When all interactions are repulsive, there
cannot be bright solitons.  However, bright solitons can be formed in
two-component repulsive BECs for a sufficiently attractive interspecies
interaction, which induces an attractive effective interaction among
bosons of same type.

\end{abstract}

\maketitle


Recently, there have been  successful observation
\cite{exp1,exp4} and associated  
theoretical \cite{yyy} 
studies of two-component  Bose-Einstein condensates
(BEC). 
There have also been experimental observation \cite{sol} of bright
solitons in a BEC 
formed due to atomic attraction and related theoretical investigations 
\cite{solt}. In fiber optics true one-dimensional
solitons are formed in the nonlinear Schr\"odinger (NLS) equation
\cite{1,agra}. The
solitons of BEC \cite{sol} are formed in a quasi-one-dimensional geometry
achieved by
employing a strong transverse trap. In either case, no solitons can be
formed for repulsive interactions.

In this Letter we suggest the
possibility of the formation of stable bright solitons  in
two-component BECs in the
presence of repulsive interaction among like atoms and attractive
interaction among atoms of different types. 
We find that  a sufficiently strong interspecies  attraction  can
induce an effective 
attraction among like atoms responsible for the formation of bright
solitons in two-component BECs. We also consider similar two-component
BECs in the presence of a periodic  optical-lattice potential, which are
now routinely used in experiments on BECs \cite{ol}.  We consider 
the formation of these   solitons 
using a coupled time-dependent mean-field 
Gross-Pitaevskii (GP)
equation \cite{11}. 

Because of two components such
solitons of the coupled one-dimensional NLS equation 
are often termed vector solitons \cite{agra} in nonlinear fiber
optics, where two 
orthogonally-polarized pulses of different widths and peak powers
in general propagate undistorted. The vector solitons considered so far in
nonlinear
optics were always created in focusing (attractive) medium. To the
best of our knowledge the present study is the first to consider the
possibility of creation of
vector solitons in defocusing (repulsive) medium.

The experimental study of bright solitons in quasi-one-dimensional
attractive systems
is quite delicate due to the possibility of collapse in such systems, 
although a true one-dimensional system does not exhibit  collapse
\cite{11}. The
two-component
repulsive BECs with interspecies  attraction are better suited for
studying solitons as such systems may not easily  collapse \cite{ska} 
and one can have a
controlled study of solitons.
Experimentally, this could be realized by
forming a coupled repulsive BEC in a cigar-shaped geometry and then
transforming the  interspecies repulsion to attraction via a Feshbach
resonance  \cite{fs} and eventually removing the axial trap so that the
BEC
components 
attain mobility in the axial direction like soliton under the action of
radial trapping alone.

Bright solitons are really eigenfunctions of the one-dimensional NLS 
equation. However, the experimental realization of bright
solitons in trapped attractive cigar-shaped BECs has been possible under 
strong transverse binding which, in the case of weak or no axial binding,  
simulates the ideal one-dimensional situation for the formation of bright
solitons. 
The 
dimensionless NLS  equation in the attractive
or
focusing case  \cite{1}
\begin{equation}\label{nls}
i u_t+u_{xx}+  |u|^2u=0,
\end{equation}
sustains the following bright
 soliton \cite{1}:
\begin{eqnarray}\label{DS}
u(x,t)&=& \sqrt{2 w}\hskip 3pt \mbox{sech}
[\sqrt{w}(x-\delta+2v t)] \nonumber \\ &\times& 
\exp[-iv(x-\delta) +i(w -v v)t+i\sigma], 
\end{eqnarray}
with four parameters. The parameter  $w$ represents the amplitude as well
as pulse width, $v$ represents velocity, the parameters $\delta$ and 
$\sigma$ are phase constants. The bright soliton 
profile  
is easily recognized for $v=\delta=0$ when Eq. (\ref{nls}) leads to
$|u(x,t)|=\sqrt{2w}\hskip 3pt  \mbox{sech} (x\sqrt{w})$.  
In the case of the coupled focusing NLS
equations:
\begin{eqnarray}
iu_t+u_{xx}+(|u|^2 +|v|^2)u=0, \label{x1}\\
iv_t+v_{xx}+(|v|^2 +|u|^2)v=0, \label{x2}
\end{eqnarray}
one could have the following bright vector solitons \cite{agra,mana}: 
$u(x,t)=\cos(\theta)\sqrt{2w} \mbox{sech} (x\sqrt{w})\exp({it}),$ $
v(x,t)=\sin(\theta)\sqrt{2w}  \mbox{sech} (x\sqrt{w})\exp({it}), $ where
$\theta$
is an
arbitrary angle. Equations  (\ref{x1}) and (\ref{x2}) are incoherently
coupled as the coupling depends only on the intensities and is therefore
phase insensitive.  In this Letter we consider the vector solitons in
Eqs. (\ref{x1}) and (\ref{x2}) with  defocusing diagonal 
nonlinearities. 
 

The time-dependent Bose-Einstein condensate wave
function $\Psi({\bf r},t)$ at position ${\bf r}$ and time $t $
may
be described by the following  mean-field nonlinear GP equation
\cite{11}
\begin{eqnarray}\label{a} \biggr[- i\hbar\frac{\partial
}{\partial t}
-\frac{\hbar^2\nabla_{\bf r}^2   }{2m}
+ V({\bf r})
+ g n
 \biggr]\Psi({\bf r},t)=0, 
\end{eqnarray}
with normalization $ \int d{\bf r} |\Psi({\bf r},t)|^2 = 1. $ 
Here 
$n\equiv  |\Psi({\bf r},t)|^2$ is the boson 
probability density,
 $g=4\pi \hbar^2 a N/m $, with
$a$ the boson-boson scattering length,
$m$
the mass and  $N$ the number of bosonic atoms in the
condensate.  
The trap potential with axial symmetry may be written as  $
V({\bf
r}) =\frac{1}{2}m \omega ^2 (\rho^2+\nu^2 z^2)$ where
 $\omega$ and $\nu \omega$ are the angular frequencies in the radial
($\rho$) and axial ($z$) directions 
with $\nu$ the anisotropy parameter, which will be taken to be 0 for
axially free solitons in the following. 

In the presence of two types of bosons each of mass $m,$
Eq. (\ref{a}) gets
changed 
to the following set of coupled equations \cite{ska}: 
\begin{eqnarray}\label{e} \biggr[- i\hbar\frac{\partial
}{\partial t}
-\frac{\hbar^2\nabla_{\bf r}^2   }{2m}
+V({\bf r})
+ g_{11} n_1+g_{12}n_2
 \biggr]\Psi_1({\bf r},t)=0,\nonumber \\ \\
  \biggr[- i\hbar\frac{\partial
}{\partial t}
-\frac{\hbar^2\nabla_{\bf r}^2   }{2m}
+V({\bf r})          
+ g_{21} n_1+g_{22}n_2
 \biggr]\Psi_2({\bf r},t)=0. \label{f}\nonumber \\
\end{eqnarray}
Here $n_i\equiv |\Psi_i({\bf r},t)|^2$, $g_{ij}
=4\pi \hbar^2 a_{ij} N_i/m,\quad $ and $   i,j=1,2$ represent the two
types of bosons, and  
where $N_1$ is the number of
boson 1 and  $N_2$ that of boson 2, $a_{ij}$ is the 
scattering 
length for a boson of type $i$ and one of type $j$.

For the study of bright  solitons
we shall reduce Eqs. (\ref{e}) and  (\ref{f}) to a minimal 
 one-dimensional  
form in the  cigar-shaped geometry with
$\nu =0$. This is achieved  by considering 
solutions of the type
$\Psi_i({\bf r},t)=  \phi_i(z,t)\psi^{(0)}( \rho)$ 
where
\begin{eqnarray}\label{wfx}
|\psi^{(0)}(\rho)|^2&\equiv&
{\frac{m\omega}{\pi\hbar}}\exp\left(-\frac{m
\omega
\rho^2}{\hbar}\right).
\end{eqnarray}
The expression (\ref{wfx})
corresponds to the
ground state wave function in the absence of nonlinear interactions and
satisfies
\begin{eqnarray}
-\frac{\hbar^2}{2m}\nabla_\rho ^2\psi^{(0)}(\rho)
+
\frac{1}{2}m\omega^2\rho^2
\psi^{(0)}(\rho)&=&\hbar\omega
\psi^{(0)}(\rho),
\end{eqnarray}
with normalization 
$2\pi \int_{0}^\infty |\psi^{(0)}(\rho)|^2 \rho d\rho=1.$
Now the dynamics is carried by $ \phi_i(z,t)$ and the radial dependence 
is
frozen in the ground state $\psi^{(0)}(\rho)$.

Averaging over the radial mode, 
i.e., multiplying
Eqs. (\ref{e}) and (\ref{f})
by  $\psi^{(0)*}(\rho)$
and integrating over $\rho$, we obtain the following one-dimensional 
 equations \cite{abdul}:
\begin{eqnarray}\label{i} \biggr[ i\hbar\frac{\partial
}{\partial t}+
\frac{\hbar^2}{2m}\frac{\partial^2}{\partial z^2}
 -f_{11}|
\phi_1|^2
-f_{12}| \phi_2|^2
 \biggr] \phi_{{1}}(z,t)=0, 
\end{eqnarray}
\begin{eqnarray}\label{j} 
\biggr[  i\hbar\frac{\partial
}{\partial t}
+\frac{\hbar^2}{2m}\frac{\partial^2}{\partial z^2}
-f_{21}|
\phi_2|^{2}  
- f_{22}| \phi_2|^2
 \biggr] \phi_{{2}}(z,t)=0, 
\end{eqnarray}
where 
\begin{eqnarray}
 f_{ij}=2\pi g_{ij}{\int_0^\infty|\psi^{(0)}|^4\rho d\rho}
=
g_{ij}{\frac{m\omega}{2\pi\hbar}}.
\end{eqnarray}
In Eqs. (\ref{i}) and (\ref{j}) 
the normalization 
is given by $\int_{-\infty}^\infty |\phi_i(z,t)|^2
dz = 1$.

For calculational purpose it is convenient to reduce 
the set  (\ref{i}) and (\ref{j})  to
dimensionless form 
by introducing convenient  dimensionless variables. 
In Eqs. (\ref{i}) and (\ref{j}) 
we consider the dimensionless variables 
$\tau=t \omega/2$,
$y=z /l$,
${\chi}_i=
\sqrt{l} \phi_i$, with $l=\sqrt{\hbar/( \omega m)}$, 
so that we have the following coupled NLS equations
\begin{eqnarray}\label{m} \biggr[  i\frac{\partial
}{\partial \tau}
+\frac{d^2}{dy^2} 
-    
 n_{11}
\left|{{\chi}_1}\right|^2        
-n_{12}
  \left|{{\chi}_2}\right|^2                  
 \biggr]{\chi}_{{1}}({y},\tau)=0,        
\end{eqnarray}
\begin{eqnarray}\label{n} \biggr[   i\frac{\partial
}{\partial \tau}+\frac{d^2}{dy^2} 
-
n_{21}
  \left|{{\chi}_1} \right|^2
-
n_{22}
  \left|{{\chi}_2}
\right|^{2}
 \biggr]{\chi}_{{2}}(y,\tau)=0,
\end{eqnarray}
where
$n_{ij}=4a_{ij}N_i/l, \quad i,j=1,2.$
 In Eqs.
(\ref{m}) and (\ref{n}),
the normalization condition  is given by 
$\int_{-\infty}^\infty |\chi_i(y,\tau)|^2 dy =1 .$

Equations (\ref{m}) and (\ref{n}) represent the one-dimensional limit of
the three-dimensional equation. For solitons we finally have to 
 take the nonlinearity coefficients $n_{ij} (i\ne j)$ to be
negative corresponding to attraction. These equations have analytic
solutions only under special conditions. 
  First, when $n_{12}=n_{21}=0$, they become two
uncoupled NLS equations which allow the
following trivial 
soliton solutions
for negative $n_{ii}$ \cite{1}:
\begin{eqnarray}\label{z1} 
\chi_i(y,\tau)= \sqrt{(2 w_i/|n_{ii}|)} \mbox{sech}(y\sqrt{w_i}) 
\exp(i\tau), 
\end{eqnarray}
$i=1,2.$ To satisfy the normalization condition one should have
$w_i=n_{ii}^2/16 $.  
In Eq. (\ref{z1}) and below the parameter $w$ is to be adjusted so as to
satisfy the normalization condition.
When all the nonlinear interactions are attractive (negative)
and $n_{11}=n_{22}=-n$ and $n_{12}=n_{21}=-\alpha n$
one has the solutions \cite{1}
\begin{eqnarray} 
\chi_i(y,\tau)= \sqrt{[2
w/(n\alpha+n)]} \mbox{sech}(y\sqrt{w}) \exp(i\tau),
\end{eqnarray} 
$i=1,2$, with $w=(n\alpha+n)^2/16$.
Also, when $n_{ii}=0$, they have the following soliton solutions for
negative nonlinearities
\begin{eqnarray} 
\chi_i(y,\tau)= \sqrt{(2w_i/|n_{ji}|)}
\mbox{sech}(y\sqrt{w_i}) \exp(i\tau), 
\end{eqnarray}
$i\ne j,$
$i,j=1,2$ with $w_i=n_{ji}^2/16$.  
This case has the possibility of forming the soliton due to interspecies
interaction as the intra-species interaction is zero.  

It is also possible to have intra-species repulsion (positive or
defocusing 
$n_{ii}$) and interspecies
attraction (negative or focusing $n_{12}$ and $n_{21}$) in order to
have the following
soliton solutions of
Eqs. (\ref{m}) and (\ref{n}) when $n_{11}=n_{22}=n$ and
$n_{12}=n_{21}=-\alpha n, (\alpha > 1):$
\begin{eqnarray} 
\chi_i(y,\tau)= \sqrt{[2w/(n\alpha-n)]}
\mbox{sech}(y\sqrt{w}) \exp(i\tau), 
\end{eqnarray} 
$i=1,2$ with $w=(n\alpha-n)^2/16$.  
In this case due to strong interspecies attraction 
solitons are formed despite of intra-species repulsion. In all these cases
the functional dependence  of the two analytic solutions 
$\chi_i(y,\tau)$
on
$y$ 
are the
same. However, there are interesting numerical solutions to
Eqs. (\ref{m}) and (\ref{n})
where the
functional dependence of $\chi_i(y,\tau), i=1,2,$ could be different, 
which
we
study next. Such vector solitons will have different widths and peak 
powers.


We solve the coupled mean-field-hydrodynamic  equations  (\ref{m}) and
(\ref{n}) for bright  solitons numerically using a time-iteration
method based on the Crank-Nicholson discretization scheme
elaborated in Ref. \cite{sk1}.  
We
discretize the mean-field-hydrodynamic  equation
using time step $0.0005$ and space step $0.025$.

\begin{figure}
 
\begin{center}
\includegraphics[width=.95\linewidth]{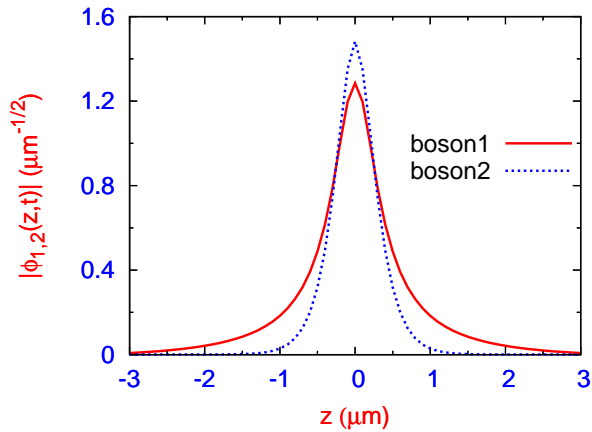}
\end{center}

\caption{(a) The stationary 
 functions $|\phi_i(z,t)|, i=1,2,$
 for bosonic  bright solitons  vs. $z$ for 
 $N_1=N_2=7500$, $a_{11}=1 $ nm,  $a_{12}=-1 $ nm, and $a_{22}=0.1$ nm,
harmonic
oscillator length $l\approx 1$ $\mu$m and $\nu =0$.  The nonlinearity
parameters are
$n_{11}=30 $, $n_{12}=n_{21} =-30$, and $n_{22}=3$.
}
\end{figure}

\begin{figure}
 
\begin{center}
\includegraphics[width=.95\linewidth]{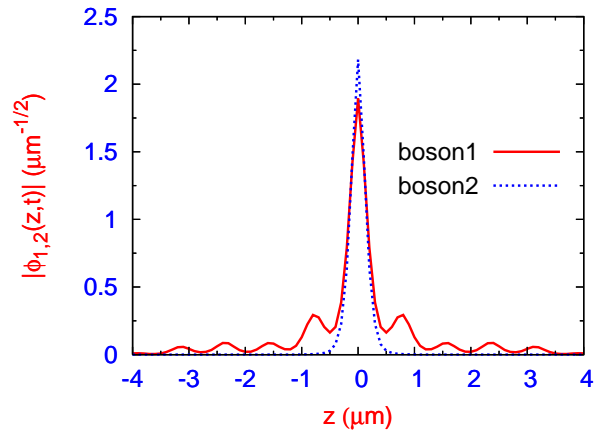}
\end{center}
\caption{(a) The stationary
 functions $|\phi_i(z,t)|, i=1,2,$
 for bosonic  bright solitons  vs. $z$ in the presence of the optical
lattice potential  $V(z)=100\sin^2 (4z)$ in Eqs. (\ref{m}) and (\ref{n}). 
 The nonlinearity
parameters are
$n_{11}=35 $, $n_{12}=n_{21} =-30$, and $n_{22}=3$.
}

\end{figure}

We performed  the time evolution of the set of equations (\ref{m}) and 
(\ref{n}) introducing harmonic oscillator potentials $y^2$ in
these equations and starting with the eigenfunction of the linear harmonic
oscillator problem with the nonlinear
terms set equal to zero: $\chi_1(y,\tau)=\chi_2(y,\tau)
=\pi^{-1/4}
\exp(-y^2/2) \exp(i\tau)$.  During
the course of time evolution the nonlinear
terms are  switched on very slowly and resultant solution iterated 
until convergence was obtained. Then the time evolution is continued
and the harmonic oscillator potential
terms ($y^2$) are  slowly switched off and the resultant solution iterated
100 000
times 
for
convergence. If converged solutions are  obtained, they correspond to the
required bright solutions.   
In the  numerical investigation we take
  $\omega = 2\pi
\times
100$ Hz, and
$m_B$ as the mass of $^{87}$Rb. Consequently, the unit of
length $l\approx 1$ $\mu$m and unit of time $2/\omega \approx 3$ ms. 

First we solve Eqs. (\ref{m}) and (\ref{n}) with  
$N_1=N_2=7500$, $a_{11}=1$ nm, $a_{12}=-1$ nm  and
$a_{22}=0.1$  
nm. 
With these
parameters the nonlinearities in Eqs. (\ref{m}) and (\ref{n}) are
$n_{11}=30 $, $n_{12}= -30$, $n_{21}= -30$, and $n_{22}=3$.

The converged  bright solitons are plotted in Fig. 1. In this case the
 function $\phi_1$ of the first component extends over a longer
region in space
compared
to the
function $\phi_2$ of the second component. It is possible to have solitons
with different 
extensions in space by varying the parameters of the system. 
The scattering length can be manipulated in the
bosonic systems 
near a Feshbach resonance  \cite{fs} by varying a background
magnetic field. By varying the scattering length and the number of
atoms we could arrive at different values of nonlinearity parameters
from those  in Fig. 1 and   thus have different extensions of the
condensates in space. 

In the situation presented in Fig. 1, in Eq. (\ref{m}) governing the
dynamics of boson 1 the two nonlinearities are $n_{11}=30$ and
$n_{12}=-30$, which may superficially indicate an effective  
nonlinearity of 0. However, the effective nonlinearity in this equation is  
 $-n_{11}|\chi_1|^2-n_{12}|\chi_2|^2$. Due to a more strongly bound
soliton of boson 2 this effective nonlinearity could become attractive
and bind the soliton of type 1. 
    
\begin{figure}
 
\begin{center}
\includegraphics[width=1.\linewidth]{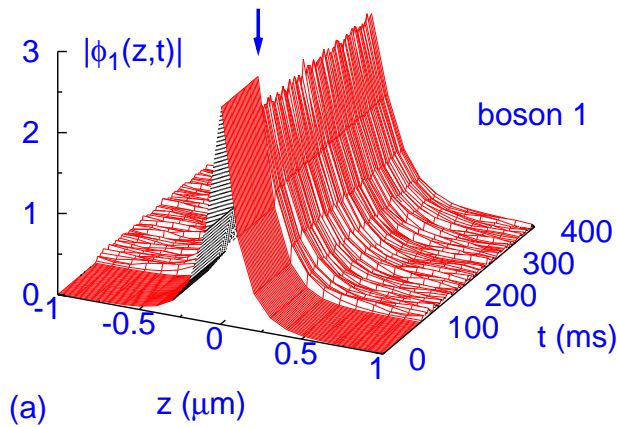}
\includegraphics[width=1.\linewidth]{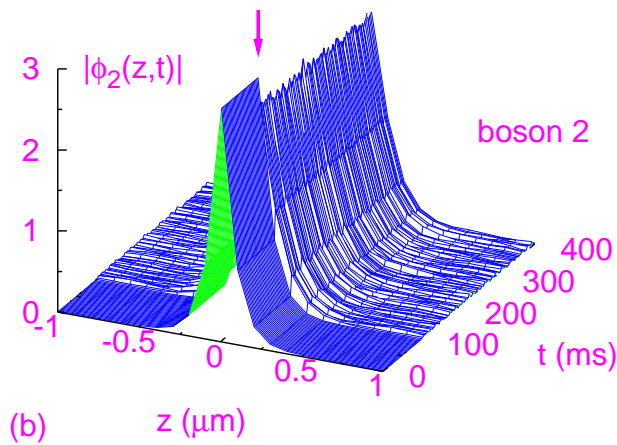}
\end{center}

\caption{The 
 function $|\phi_i(z,t)|$
 for bosonic  bright solitons  vs. $z$ and $t$ for  (a) boson 1 and
(b) boson 2. At $t=0$  
 $N_1=N_2=7500$, $a_{11}=1 $ nm,  $a_{12}=-1.7 $ nm, and $a_{22}=0.1$ nm,
harmonic
oscillator length $l\approx 1$ $\mu$m and $\nu =0$.  The nonlinearity
parameters at $t=0$ are
$n_{11}=30 $, $n_{12}=n_{21} =-51$, and $n_{22}=3$. At $t=100$ ms (marked 
by arrows)
the
two bright solitons are  set into
breathing oscillation  by suddenly jumping the nonlinearities
$n_{12}$ and
$n_{21}$ to $n_{12}=n_{21}=-45$. 
}

\end{figure}

Next we consider a two-component soliton formed in the optical-lattice
potential $V(z)=V_0\sin^2 (4z)$ introduced in Eqs. (\ref{m}) and
(\ref{n}). In our numerical calculation we take  $V_0=100$,  
$n_{11}=35 $, $n_{12}=n_{21} =-30$, and $n_{22}=3$. In this case in the
wave function of boson 1 prominent wiggles are formed due to the
optical-lattice potential. As the spacing of the optical-lattice sites are
relatively large no wiggles are formed in the more localized wave function
of boson 2. Equation (\ref{m}) taken separately has nonlinearities
$n_{11}=35 $ and  $n_{12}=-30$ corresponding to an apparent overall
repulsion.  However, the strongly bound soliton of boson 2 and the
interspecies
attraction as well as the periodic optical-lattice potential 
aid in the formation of the soliton of boson 1. In this connection it
should be noted that the optical-lattice potential aids in binding the
soliton in the presence of nonlinear interspecies attraction. The
optical-lattice potential alone cannot bind a soliton in the absence of
nonlinear attraction both in one- and two-component BECs.

Finally, we consider the stability of these solitons under a small
perturbation. For this purpose we consider the solitons of
Eqs. (\ref{m}) and (\ref{n}) formed for nonlinearities $n_{11}=-30$,
$n_{12}=n_{21}=-51$, and $n_{22}=3$. The wave functions $\phi_i(z,t)$ of
these 
solitons   are shown in Figs. 3 (a) and (b) for $i=1,2$,
respectively. To test their stability under small
perturbation, after their formation,  the nonlinearities
$n_{12}$ and $n_{21}$ are suddenly changed to $-45$  at $t=100$ ms so
that the solitons
are set into  motion. Such change in the nonlinearity can be achieved by a
jump in the scattering length by manipulating a background magnetic field
near a Feshbach resonance \cite{fs}. The solitons are found to execute
stable
non-periodic breathing 
oscillations. The stability of these solitons after the perturbation is
applied is demonstrated in Figs. 3.

\begin{figure}
 
\begin{center}
\includegraphics[width=.95\linewidth]{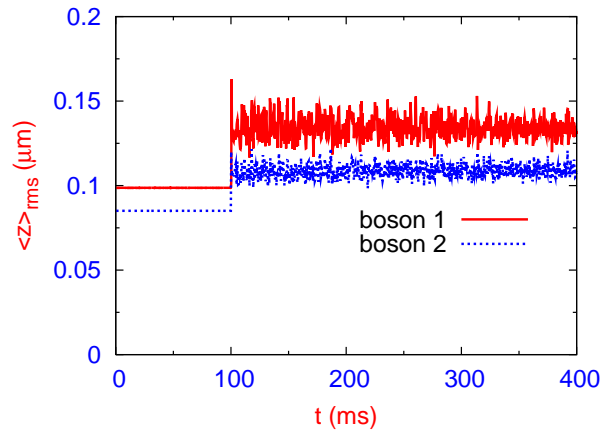}
\end{center}

\caption{Root mean square size  $\langle
z\rangle_{\mbox{rms}}$ vs. time of the two solitons of Fig. 3 set into
breathing oscillation by suddenly jumping the nonlinearities at $t=100  $
ms. 
}
\end{figure}

To further illustrate the stability of the oscillation of the 
solitons of
Figs. 3 we plot in Fig. 4 the root mean square (rms) size $\langle
z\rangle _{rms}$ vs. time.  Sustained oscillation for a very long time
illustrates the stability of the solitons. From Fig. 4 we find that for
$t<100$ ms the rms sizes  of the two solitons are constant. However, after
the application of the repulsive  impulsive force at $t=100$ ms by
reducing the attractive nonlinearity, the rms sizes suddenly jump to a
larger value and execute  stable  oscillatory dynamics.   This stable
oscillation guaratees the stationary nature  of the solitons under small
perturbation.

The present investigation has consequences in generating bright
vector solitons
in directional couplers of nonlinear fiber optics \cite{agra}. Vector
solitons are solutions of coupled one-dimensional NLS
equations with the property that the orthogonally polarized components
propagate in a birefringent fiber without change in shape. In vector
solitons an input pulse maintains not only its intensity profile
but also its state of polarization even when it is not launched along one
of the principal axes of the fiber. We have investigated the possibility
that two orthogonally polarized pulses of different widths and different
peak powers propagate undistorted in birefringent fibers. 
The present investigation suggests that it is 
possible to have bright vector solitons in the coupled NLS equation with
diagonal defocusing (repulsive)
nonlinearity and off-diagonal focusing (attractive) nonlinearity,
which should be of interest in nonlinear fiber optics.

 
We use a coupled set of time-dependent mean-field GP equations for a
two-component repulsive BEC to demonstrate the formation of bright
solitons due to interspecies attraction. The interspecies attraction can
neutralize the intra-species repulsion and induce an effective attraction
in the mean-field GP equations responsible for the formation of bright
solitons.  An attractive interspecies interaction is necessary for the
formation of the bright solitons as the diagonal nonlinearity $n_{ii}$ in
the mean-field equations is taken to be repulsive (positive). In
mean-field equations (\ref{m}) and (\ref{n}) $n_{ii}$s are positive
(repulsive) and $n_{ij}$s are negative (attractive), so that the overall
contribution of the nonlinear terms in these equations become attractive
to support the bright solitons.  We have also 
established the formation of these solitons in the presence of a periodic
optical-lattice potential in an entirely different shape and trapping
condition from a conventional soliton.

In view of the present study the
appearance of bright solitons in multi-component repulsive BECs seems
possible in  quasi-one-dimensional
geometry.  Bright solitons have been created
experimentally in attractive BECs in three dimensions in the presence of
radial trapping only without any axial trapping \cite{sol}. Also, there
have been experimental studies of multicomponent BECs \cite{exp1,exp4}.
Hence,
bright
solitons can be created and studied  in the laboratory in the presence of
radial
trapping only in a two-component repulsive BEC supported by interspecies
attraction and the prediction of the present study verified. 
We have also suggested the
possibility of the formation of similar coupled solitons in fiber optics
in one dimension.

Note added in proof: After the completion of this investigation we have 
known about another 
similar study \cite{sim}.
 
\acknowledgments

The work is supported in part by the CNPq 
of Brazil.


\end{document}